# Analysis of Local Samples of Paracetamol at Bamako by Reflectance Near-Infrared Spectroscopy

**Aminata Sow[1, *], Issiaka Traore[1], Tidiane Diallo[2, 3], Abdramane Ba[1]**

[1]Departmenf of Physics, Faculty of Science and Technique, Laboratory of Optics Spectroscopy and Atmospheric Science, University of Sciences, Techniques and Technologies of Bamako, Bamako, Mali

[2]Department of Medicine Sciences, Faculty of de Pharmacy, University of Sciences, Techniques and Technologies of Bamako, Bamako, Mali

[3]National Laboratory of Health, Bamako, Mali

**Email address:**
aminasow100@gmail.com (Aminata Sow), aminata.sow@usttb.edu.ml (Aminata Sow)
[*]Corresponding author

**Abstract:** The approach of Near-Infrared Spectroscopy (NIRS) together with Chemometric techniques are used in order to analyze sixty (60) tablets of paracetamol of different batch numbers in the local markets in Bamako. The primary objective is to model these samples by doing multivariate regression computation. Prior to this, various statistical pretreatment methods such as second derivative (SD) correction, first derivative correction (FD), multiple scattering correction (MSC), smoothing the spectra (smooth), the standard variate normalization (SNV) correction and some combinations are performed. The Partial least square (PLS) regression on the SNV data preprocessing resulted in the detection of two outliers. Additionally, the presence of nonlinear effects is conducted. Its presence compels one to consider nonlinear regression such as the artificial neural network ANN or relevance vector machine RVM. A supporting fact of the use of these types of regressions is that ANN regression applied to the spectra under consideration significantly improves the RMSEP and the relative standard error of prediction RSEP. To further analyze the samples, the selection of wavelengths based on the p-value approach proved its usefulness in this investigation. The best calibration of the PLS multivariate regression model is obtained with the MSC combined with FD correction, and its statistical values for the fourteen wavelengths, having the smallest individual p-value, are $R^2$= 85.26%, RMSEP= $2.38 \times 10^{-4}$ and finally RSEP=1.45%.

**Keywords:** Paracetamol, Near-Infrared Spectroscopy, Chemometric Techniques, Partial Least Squares Regression, Data Preprocessing, Variable Selection, Nonlinearity

## 1. Introduction

Counterfeit pharmaceutical medicines constitute a substantial danger to the health of patients and a great financial loss for the pharmaceutical industries. Hence, the fight against falsified drugs is of capital interest worldwide. The first stage of this battle is surely developing possibilities to detect fraudulent drugs. Several mechanisms in this direction have been put forwards in the past. However, most of them are very expensive, extremely time-consuming and require specific expertise. An additional drawback of these classical models is the requirement of dissolving the samples into some other chemical solvent. As a result of this action, the samples are totally or partially destroyed during the process of identification. Nevertheless, techniques that overcome these negative aspects have been around for a while and they should be investigated seriously in every country. One method which is currently much in the air is the application of Near-Infrared Spectroscopy (NIRS) together with Chemometric techniques. This non-destructive method has been indicated as a potential alternative to surmount limitations related to sample dissolution processes.

The application of NIRS methods to analyze data, especially paracetamol samples, has yielded significant results in the past [1-6]. However, before its application, the data must undergo some preprocessing so that to eliminate some unwanted errors coming from the many steps in the measurement process. In this investigation, preprocessing methods such as second derivative (SD), first derivative correction (FD) [7] by the method of Savitsky and Golay, multiple scattering correction (MSC) [8], smoothing the spectra (smooth) [9], the standard variate normalization (SNV) [10] and naturally some combinations of them are applied. These pretreatments undoubtedly eliminate the multiplicative as well as the additive noises in the original spectra. These preprocessing steps in the application of NIRS are very essential and one has to do them. Yet these are not the end of the problems. In the calibration analysis, one may face some other types of issues such as nonlinearity or selecting the best subset amongst the predictive variables. These two problems are not new in the field of NIR spectroscopy.

Firstly, for detecting nonlinearity in NIRS spectra, the method adopted here is based on the plot of the PLS Y-score against its X-score [11, 12]. Other algorithms [13, 14] can be used as alternative directions for the same purpose. Nonlinearity in the spectra can indeed have many sources [12], but in this investigation, one is mainly concerned with detecting them. Secondly, it is important to stress that in multivariate calibration for better prediction of output variables, one may need to identify the best subset of variables from the wavelengths used in the measurement. To this end, the selection using the individual p-value approach known as backward selection [15, 16] and the one advocated by the authors of [17] are applied. The statistical p-value parameter is very important in the sense that it is related to the null hypothesis [18]. The null hypothesis is the statement that the set of instrumental (the absorbance) measurements [X] obtained at the various wavelengths have no relationship of whatsoever with the response variable which is the content [Y] of the paracetamol samples. Samples having the smallest p-values are the ones that contribute a lot to the multivariate regression algorithm. In this article, 10 and 14 wavelengths where the PLS regression has a smaller p-value are analyzed. In addition to the PLS regression, Artificial Neural Network ANN, and Relevance Vector machine RVM regressions are applied to these two subsets. The comparison between these two nonlinear regressions and PLS is also reported as far as the RMSEP and RSEP of the two subsets are concerned. Additionally, outliers may also be part of the game. The existence of this last troublemaker is something very much debated in the field of NIR spectroscopy, see for instance [18-22]. The removal of the outliers dramatically reduces the RMSEP; hence, improving the calibration performance thereof. The detection of them based on the Residuals vs Leverage or using the graph of Cook's distance is adopted in this manuscript. This approach is very well developed [23]. The overall purpose of this article is to set up an immediate and imperative program for mastering techniques used in near-infrared spectroscopy when dealing with drugs. In this program, linear and nonlinear regressions along with classification are the main targets. In this manuscript, linear regression PLS, wavelength selection to improve the results, outliers' detection and finally two nonlinear regressions (artificial neural network and relevance vector machine) applied to some of the selected wavelengths are the main debated topics.

The remaining parts of this manuscript are divided as follows: in section two, the materials and the methodologies used for spectral acquisitions are presented. Section three is devoted to the different results obtained in this manuscript. The next section deals with the discussion of those results. The conclusion and perspective are the heart of section five. This article ended up acknowledging the many supports received upon working on this project.

## 2. Materials and Methodologies

### 2.1. Materials and Data Acquisition

#### 2.1.1. Samples and Samples Preparation

In this report, sixty paracetamol tablets having different batches numbers are used. They are collected from different pharmacies in the district of Bamako to cover entirely the capital of Mali. This is very important in the sense that the remaining localities in Mali get their drugs from the capital. The sixty samples were randomly selected from 30 pharmacies in the district of Bamako. Two samples of paracetamol were collected in each pharmacy. However, in the process of sampling, a given batch number is selected only once to avoid duplication. Before doing the calibration, two samples were identified as outliers and are then removed thereof. The remaining fifty-eight different batch numbers of paracetamol are thus considered in the calibration process. The limitation to this number is not a problem in itself since in the literature fewer numbers of paracetamol have been analyzed by the techniques applied here.

In the second part of this work, a solution of NaOH with a concentration of 0.1N is used for the determination of the optical density which is used in the computation of the content of the samples under consideration.

#### 2.1.2. NIR Data Acquisition

In performing the spectral reflectance measurements an optical flame-NIR-INTSMAS25 (800- 1800 nm) spectrophotometer connected to a diffuse reflectance probe is used. To acquire the data, the optical fiber is linked to the source incorporated in the probe which is connected to a computer via the spectrometer. To measure the near Infrared spectra of any sample, the subjected samples are firstly labeled, triturated and passed through a sieve with a diameter of 250 μm. A weighed of 0.206 g is taken from the obtained powder so that to uniformly have the same quantity, hence the same surface area in the measurement process. Since the ambient light is also a source of problems in the process of measurement, the experimental facilities are set to an extent such that the complete system (apparatus + sample) can be considered as being hermetic as possible as it should be. This has the advantage that when passing light onto the sample on the plate of the analysis device, scattering effects are much more reduced. As for the next step in the acquisition, the light of the spectrometer is projected on the samples paying due attention to the fact that at every time there is no penetration of light from outside. Finally, in proceeding with recording the spectra, the samples are scanned 10 times each sample the spectrum of a given sample is the average of these results. These different steps have been pursued exactly for every sample in this investigation. The results yield the data matrix $X_{58*128}$ of 58 rows representing the different samples and 128 columns denoting the wavelengths at which the spectra have been recorded. This spectral data matrix thus obtained is composed of the reflectance of the different paracetamol which needs to be transformed into absorbance.

To perform the multiple variate regression, one needs to determine the contents of the paracetamol samples which is done by using a UV-Visible spectrometer.

### 2.1.3. UV-VISIBLE Data Acquisition

To carry out his part, the UV-visible spectrometer (Agilent Carry 630) is used. The first step leads to obtaining the optical density of the drug which is used in the computation of the content of the paracetamol. To successfully get optical density; hence, the content, one closely follows the labeling procedure as mentioned above. However, one instead took the average weight of 20 tablets of each sample. The following step is the computation of the test portion (TP), weighing the powder of each sample and dissolving it into NaOH solution at a concentration of (0, 01N). Finally, one proceeds with the data acquisition by scanning 10 times the prepared overall solution and taking their averages. This average corresponds to the Optical density (OD) which yields a data matrix of 60 rows for the samples and 2 columns for Optical Density and content respectively, we will denote it by $D(58 \times 2)$. The content is denoted by $T(58 \times 1)$.

### 2.2. Data Preprocessing

With the data matrices obtained in NIRS and UV-Visible measurements, one built a combined data consisting of 58 rows for the samples and 129 columns for the wavelength and the content. To build a reliable predictive model, one must then pretreat the obtained matrix. This is primarily because many undesirable factors bring fluctuations in the measurement process. Consequently, to handle these unwanted variations the data in this report goes under various pretreatments techniques. This section deals with the aforementioned pretreatments in the abstract. To do so, one relies on the references mentioned in the introduction. In general, the spectrum $A(\lambda)$ measures at the wavelength $\lambda$ can be decomposed as follows:

$$A(\lambda) = \alpha . A_0(\lambda) + \beta + \varepsilon(\lambda) \qquad (1)$$

where α is a multiplication scatter factor; $A_0(\lambda)$ is the true real spectrum one is looking for; β is an additive scatter factor also known as the offset deviation and $\varepsilon(\lambda)$ is the wavelength-dependent noise. The central theme of spectral pretreatments is to find ways of reducing the effects of these three factors. The simultaneous application of many different pretreatment techniques usually yields different results. Therefore, it is advisable to apply many of them and seek possible improvements in model building. In the present work, the measurement at different wavelengths of the Reflectance of the different paracetamol is been done in reflectance mode and the result matrix is denoted by $R(I \times J), I = 1, 2, 3, \ldots, 58$ and $J = 1, 2, 3, \ldots, 128$.

The first entry I stands for the ith samples and the second entry J denotes the jth wavelength. However, all the computations are performed by using the Absorbance data matrix, which is obtained by using

$$A = A_{I,J} = \log_{10}\left(\frac{1}{R}\right) \qquad (2)$$

Before any pretreatment technique, one firstly auto-scales the matrix of absorbance by mean centering and variance scaling it. This process is very standard therefore it is not repeated here.

To this final matrix, the aforementioned pretreatment methods are applied. These pre-processing techniques are split into two groups: the ones which correct the multiplicative scatter factor and the last group that redresses the derivatives behaviors. Two methods of the first group (multiplicative scatter correction-MSC and standard normal variate-SNV) are first applied. It is known that these two types of pretreatments are linearly related [22]. Next, three other corrections from the second group (the first, second derivative and the Savitzky-Golay smoothing technique) are then applied. The mathematical algorithms of these techniques are very well known in the literature, one can consult for instance [23-25]. They are performed with the software RStudio 2022.02.2 Build 485 and the different graphs are available [30].

### 2.3. Wavelengths Selection

Wavelength selection is a challenging problem in NIR spectroscopy. In this article, three wavelength regions are selected based on the PLS loading vectors of the principal component approach [17, 32], the p-value approach and the wavelengths of the peaks of the second derivative. These bands are characterizing the samples of paracetamol.

### 2.4. Outliers' Detection

To detect the outliers, Cook's distance of the linear model between the predicted content from the PLS and the observed content has been used. The outcome is the detection of three samples as outliers. However, one sample is kept in the model building.

## 3. Results

### 3.1. Contents

The relative information about the content of the samples are given below in table 1.

*Table 1. The results of the measurements done by UV-visible.*

| Denomination | Maximum | Minimum | Specification (%) |
|---|---|---|---|
| OD (Ech) | 0.73507 | 0.673565 | |
| Content | 105.026129 | 95.6028359 | 95-105 |
| OD (Std) | 0.7105625 | | |
| PM | 593.05 | 547.42 | |
| TP | 118.61 | 109.484 | |

Where OD (Ech) is the optical density of the sample, OD (Std) is the optical density of the standard sample of paracetamol, PM (mg) is the mean of the weight and TP (mg) is the test portion. The standard formula for the content is given by

$$T(\%) = DO(Ech) * 100 * 100 * dilution(0.01) * \frac{PM}{DO(std)} * TP * \frac{100}{500} \qquad (3)$$

### 3.2. PLS Loadings

The graph of the loading from the PLS is given by figure 1.

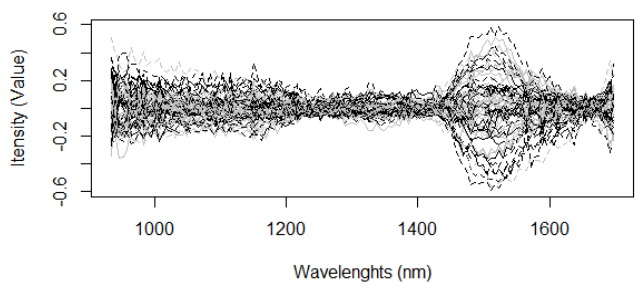

*Figure 1.* PLS loading vectors of the spectra.

This graph is used in choosing the band of the loading plot.

### 3.3. The Graph of the Second Derivative

To localize the peaks of the absorbance, one uses the derivative pretreatments. Namely, the graph of the second derivative. The following plots show the second derivative and the peaks obtained.

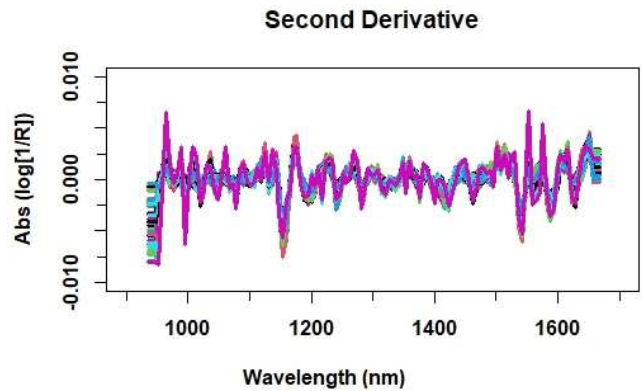

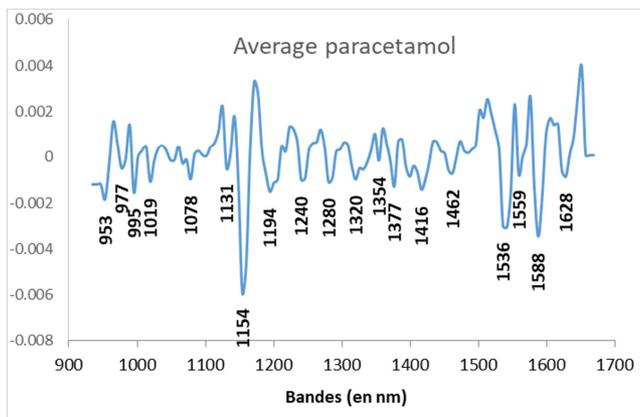

*Figure 2.* Second derivative and the peaks (bands or wavelengths) associated to it.

One may then compare these peaks to what is available in the literature. This is the subject of future works.

### 3.4. Detection of the Outliers

The next graphs are the ones for the detection of outliers and they are presented below.

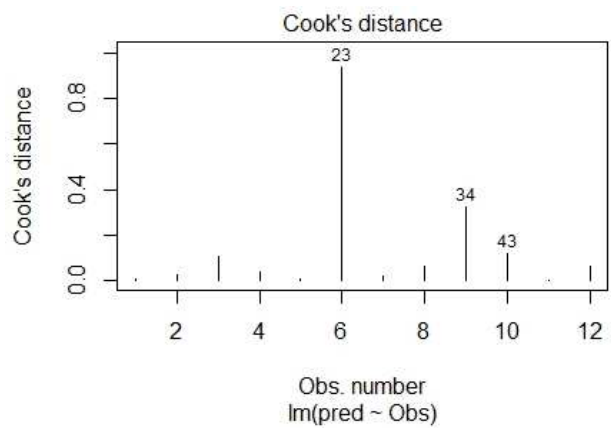

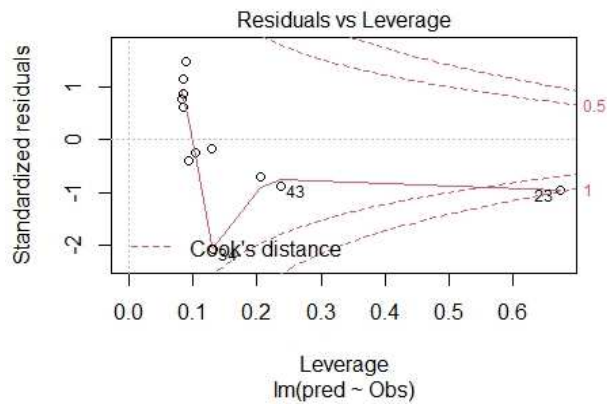

*Figure 3.* Detection of the outliers, samples 23 and 34 are removed from the investigation. Pred is the predicted contents by PLS and Obs is the measured content by UV-Visible.

In the first PLS model, 48 samples have been selected to form the calibration set and the remaining 12 samples were then chosen as the validation set or test set. Samples 23, 34, and 43 were part of the set. The calibration result was terribly bad signaling the presence of outliers therein lie the problem. To handle them, an approach based on Cook's distance has been applied to the PLS-predicted results of the contents. The result of this detection is shown in the plot above. The Cooke distance of sample 23 is very large, then follows that of 34. Therefore, these two samples are removed in what to follows.

### 3.5. Detection of Nonlinearity

The presence of nonlinearity can lead to fewer problems in calibration. Hence, a precise mechanism must be applied for detecting it. The following graph showed the plot of the Yscore again the Xscore of the PLS model for the first three components.

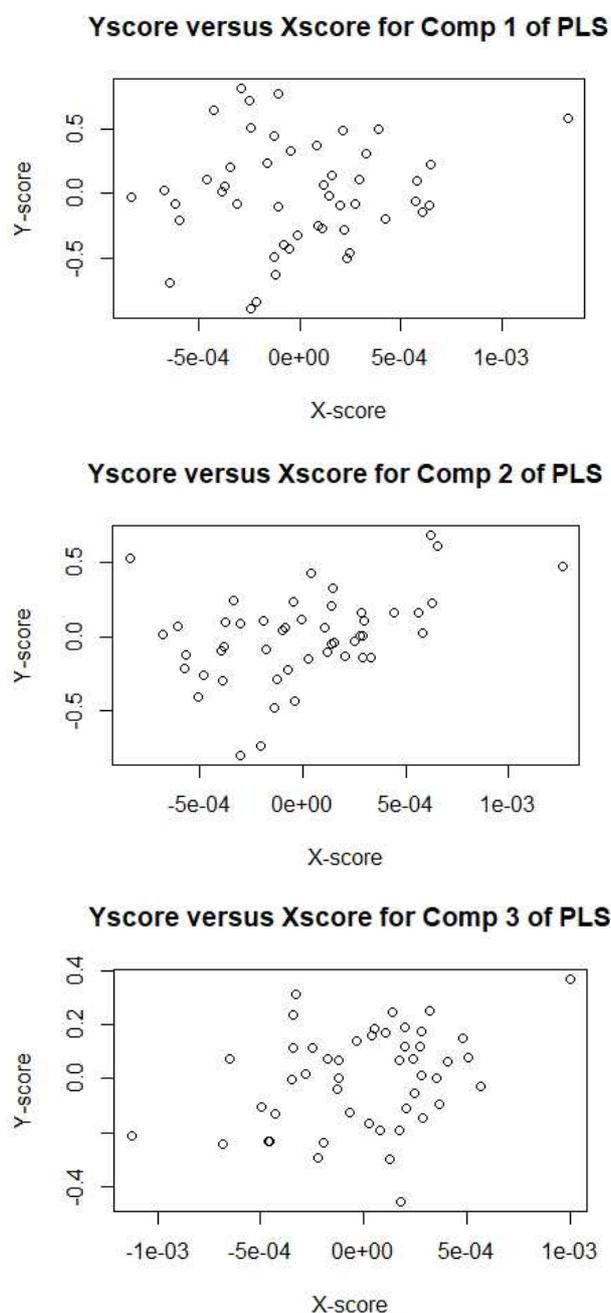

*Figure 4. Detection of nonlinearity (component (comp) 1, 2 and 3).*

This detection is a signal that nonlinear regression techniques are better adapted to the spectra of paracetamol considered in this investigation. Nonlinearity can sometimes seriously impact the outcome of a regression model. It is therefore mandatory to deal with them in a manner that reduces the drawbacks coming from them. Nonlinear regressions are designed to do so. Hence, they are perfect for the samples used in this investigation. This is supported by the results of [30].

### 3.6. Construction of the PLS Models

Several multivariate regression algorithms are available for analyzing NIR data. In this work; however, the partial least square model (PLS) regression is adopted. This is firstly due to its capacity to easily detect structure within the matrix of absorbance data which has predictive relevance for the content of the samples and secondly it can be used to classify new samples with the only knowledge of the absorbance data of that sample. A pedagogical overview of this technique is given [26]. To do PLS, one has to partition the samples into a train set and a test set. The former is used for model building and the latter is designed for testing the model. One fundamental problem is then to decide which samples should be in the train set or test set. Random selection, or using a few algorithms for the purpose that are already available in the literature, can partially overcome this issue. In the present model, this problem is bypassed by making the combination of many techniques and selecting 48 samples in the train set and 12 samples for the test set. However, after detecting the 2 outliers, the remaining 10 samples were employed as the test set in the article.

This is not the end of history yet. An additional challenge facing any PLS model is the wavelength selection procedure. In this present manuscript, the first selected region is naturally the full range of the wavelength at which the reflectance of the paracetamol has been measured. The next, selected region is the peaks that one gets from the loading of PLS regression as advocated [17]. Another region is the band from 1405 nm up to 1633 nm (Figure 1) which shows noticeable variability in the plot of the SNV. Last but not least is the selection from the smallest p-value in the spirit [15].

To begin with the computation of the parameters of the PLS regression model, it should be stressed that the algorithm seeks a prediction of the content $T_p[i] (i = 1, 2, \ldots, 58)$ of the samples in terms of the values of the pretreatments collectively denoted by $X[i,j], (i = 1, 2, \ldots, 58 \text{ and } j = 1, 2, \ldots, w = 128)$ at wavelength j, where w stands for the range of the selected region of wavelength. The equation relating these two quantities is given by

$$T_p[i] = T_0 + \sum_{j=1}^{P} B_j X[i,j] \qquad (4)$$

The PLS algorithm statistically yields the parameters of $T_0$ and $B_j$ in equation (4) and one can simply use them in predicting the content of new samples. One looks at the following statistical parameters: the multiple correlation coefficient $R^2$, the Root Mean Square Error $RMSE$, Mean Absolute Error MAE, and the number of PLS components used by (4) which was selected to minimize the residual error sum squares for the validation set as advocated [2]. These parameters are given by

$$R^2 = 1 - \left( \frac{\sum_{i=1}^{N}(T_p[i] - T[i])^2}{\sqrt{\sum_{i=1}^{N}(T[i] - \bar{T})^2}} \right) \qquad (5)$$

$$RMSE = \sqrt{\frac{\sum_{i=1}^{10}(T_p[i]-T[i])^2}{10}} \qquad (6)$$

$$MAE = \frac{\sum_{i=1}^{10}|T_p[i]-T[i]|}{10} \qquad (7)$$

Where $T[i], \bar{T}$ and $T_P[i]$ represents the observed, its mean, and the predicted value of the content, respectively. The results of the computation are given in Table 2 below. As one can see, five pretreatments single themselves out as the best models which are SNV at full wavelength range, simple without no preprocessing in the range 1405 nm-1633 nm, the SNV for the 12 wavelengths characteristic of paracetamol and naturally the models based on the p-values. In Table 2, MSC+FD is the MSC correction followed by the first derivative. Meanwhile, SNV+SD is the application of the second derivative to the correction by SNV. Smoothing is the result obtained by the application of Savitzky-Golay smoothing correction. The No. PLS-comp (the number of data points that have Leverage) can be obtained by plotting the Residuals vs Leverage. Statistically, they influence the regression results; hence, excluding them will strongly alter the regression results. The result of the different models constructed can be compared based on the value of the relative standard error of prediction from table 3.

*Table 2.* PLS-MODELS RESULTS.

| | Treatment | RMSE ($10^{-4}$) | $R^2$ (%) | MAE ($10^{-4}$) | No. PLS-comp. |
|---|---|---|---|---|---|
| Full wavelength | SIMPLE | 2.54 | 77.52 | 2.13 | 4 |
| | SNV | 3.07 | 87.80 | 2.44 | 3 |
| | MSC | 3.24 | 78.30 | 2.50 | 3 |
| | FD | 3.15 | 71.80 | 2.40 | 1 |
| | SD | 3.55 | 38.55 | 2.65 | |
| | MSC+FD | 3.06 | 74.52 | 2.39 | 2 |
| | MSC+SD | 3.61 | 60.71 | 3.13 | 2 |
| | SNV+SD | 2.76 | 68.56 | 2.01 | 5 |
| | Smoothing | 2.48 | 72.59 | 2.03 | 5 |
| Bands: 1405nm-1633nm (41 wavelength) | SIMPLE | 3.05 | 80.83 | 2.49 | 4 |
| | SNV | 3.33 | 73.28 | 2.81 | 3 |
| | MSC | 3.37 | 69.91 | 2.84 | 3 |
| | FD | 3.15 | 62.39 | 2.25 | 3 |
| | SD | 3.84 | 53.06 | 2.80 | 9 |
| | MSC+FD | 2.81 | 76.95 | 2.18 | 2 |
| | MSC+SD | 2.95 | 56.80 | 2.15 | 8 |
| | SNV+SD | 2.96 | 56.33 | 2.16 | 8 |
| | Smoothing | 3.09 | 70.23 | 2.24 | 1 |
| Bands: 952nm, 958nm, 983nm, 1007nm, 1461nm, 1490-1518nm (6), 1547nm (12 bands) | SIMPLE | 3.08 | 65.39 | 2.24 | 2 |
| | SNV | 2.99 | 89.05 | 2.10 | 3 |
| | MSC | 3.10 | 63.99 | 2.26 | 1 |
| | FD | 2.86 | 70.23 | 2.13 | 3 |
| | SD | 3.09 | 50.37 | 2.15 | 4 |
| | MSC+FD | 3.54 | 67.43 | 2.65 | 2 |
| | MSC+SD | 3.08 | 53.94 | 2.25 | 4 |
| | SNV+SD | 3.42 | 75.90 | 2.51 | 5 |
| | Smoothing | 3.38 | 62.88 | 2.46 | 2 |
| 10 wavelengths Having smallest Individual P-Value | SIMPLE | 3.48 | 75.69 | 2.50 | 2 |
| | SNV | 3.09 | 58.75 | 2.14 | 3 |
| | MSC | 3.61 | 66.62 | 2.59 | 2 |
| | FD | 2.61 | 81.04 | 1.85 | 2 |
| | SD | 3.57 | 40.59 | 2.82 | 3 |
| | MSC+FD | 2.46 | 72.15 | 1.89 | 2 |
| | MSC+SD | 3.44 | 64.54 | 2.59 | 4 |
| | Smoothing | 3.04 | 68.69 | 2.21 | 2 |
| 14 wavelengths Having smallest Individual P-Value | SIMPLE | 4.29 | 62.75 | 3.35 | 4 |
| | SNV | 3.47 | 50.76 | 2.54 | 9 |
| | MSC | 3.60 | 53.18 | 2.75 | 2 |
| | FD | 2.67 | 81.16 | 1.93 | 2 |
| | SD | 3.11 | 62.60 | 2.28 | 8 |
| | MSC+FD | 2.39 | 85.26 | 1.79 | 2 |
| | MSC+SD | 3.05 | 71.07 | 2.19 | 7 |
| | Smoothing | 3.07 | 60.61 | 2.24 | 3 |

To further decide on the best regression model, the following parameters of the different models are computed. They are the Mean bias, the Mean accuracy and the Relative Standard Error of Prediction RSEP given [2] (n=10) as follows:

$$Mean\ bias\ (\%) = \left[\sum_{i=1}^{10} \frac{T_p[i]-T[i]}{T[i]} \times 10\right] \qquad (8)$$

$$\text{Mean accuracy (\%)} = \left[\sum_{i=1}^{10} \frac{|T_p[i]-T[i]|}{T[i]} \times 10\right] \quad (9)$$

$$RSEP(\%) = \sqrt{\frac{\sum_{i=1}^{10}(T_p[i]-T[i])^2}{\sum_{i=1}^{10}(T[i])^2}} \times 100 \quad (10)$$

The results of these computations are given in Table 3. The residual error of prediction in Table 3 is the parameter that will be used in deciding upon the best regression model. The smaller is RSEP, the better the model is.

*Table 3. FURTHER PLS-MODELS QUANTIFICATION.*

| Bandes | Treatment | Bias ($10^{-4}$) | Mean Bias (%) | Mean Accuracy (%) | RSEP (%) |
|---|---|---|---|---|---|
| Full Spectrum | SIMPLE | -2.08 | 1.27 | 1.31 | 1.54 |
| | SNV | -2.44 | 1.5 | 1.5 | 1.86 |
| | FD | -2.29 | 1.41 | 1.48 | 1.91 |
| | MSC | -2.5 | 1.54 | 1.54 | 1.97 |
| | MSC+FD | -2.3 | 1.42 | 1.47 | 1.85 |
| Bands: 1405nm-1633nm (41 wavelength) | SIMPLE | -2.49 | 1.53 | 1.53 | 1.85 |
| Bands: 952nm, 958nm, 983nm, 1007nm, 1461nm, 1490-1518nm (6), 1547nm | SNV | -2.09 | 1.29 | 1.29 | 1.81 |
| 10 wavelengths Having smallest Individual P-Value | FD | -1.56 | 0.97 | 1.14 | 1.58 |
| 14 wavelengths Having smallest Individual P-Value | MSC+ FD | -1.43 | 0.88 | 1.10 | 1.45 |

In Table 4, the comparison between the best PLS model and the result from ANN and RVM is presented. This is used to merely appreciate the best model of this manuscript. These four tables enable one to conclude with confidence which PLS model is the best regression model for analyzing the present paracetamol samples by near-infrared spectroscopy. Table 4 also indicates that nonlinear regression techniques (which are beyond the scope of this manuscript) should be applied to the present samples of paracetamol. These applications are performed elsewhere [30].

*Table 4. Best PLS Compared to ANN and RVM.*

| Regression | Bands | Preprocessing | RSEP (%) |
|---|---|---|---|
| PLS | 10 wavelengths Having the smallest Individual P-Value | FD | 1.58 |
| | 14 wavelengths Having the smallest Individual P-Value | MSC+FD | 1.45 |
| ANN | 10 wavelengths Having the smallest Individual P-Value | FD | 1.00 |
| | 14 wavelengths Having the smallest Individual P-Value | MSC+FD | 1.12 |
| RVM | 10 wavelengths Having the smallest Individual P-Value | FD | 0.07 |
| | 14 wavelengths Having the smallest Individual P-Value | MSC+FD | 0.16 |

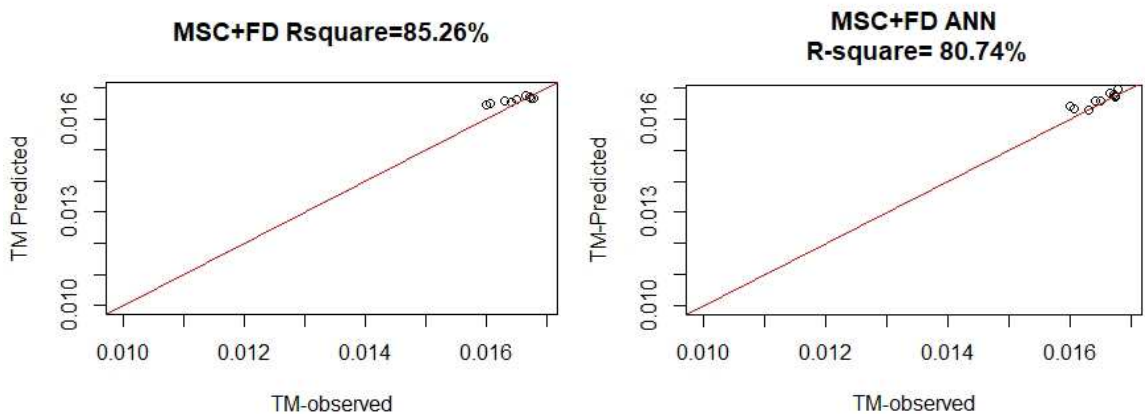

*Figure 5. The predicted values against the observed values of the contents.*

The following two graphs represent the plot of the prediction against the observed value of the contents with MSC+FD corrections for PLS and ANN.

## 4. Discussions

Having presented the different results obtained in this work, it is now the right place to discuss the importance of some of the results. First of all, after a close scrutinizing of the plots, one notices that three samples are outliers, but two samples are removed in this investigation. What is more, the plot of the second derivative is highlighting useful information about the peak associated with the different excipients of the paracetamol. These peaks can also be seen as associated with the different atomic liaisons in the formula of paracetamol. As such, they can be compared to the peaks of numerous overtone and combination bands found for instance [27, 28]. One may also use the graph of the second derivative and compare it to the one from adulterated paracetamol. This usage of the plot of the second derivative correction is adopted [31].

The removal of two of them significantly improved the PLS regression results. Next, one notices that the p-value approach for selecting wavelengths is a valuable procedure. Moreover, the importance of the different preprocessing techniques is manifestly seeable from the different tables.

In table 1, one notices that the contents of the paracetamol are in the range of what is standard. From table 2, one sees that the PLS of the MSC followed by FD for the 14 wavelengths having the smallest Individual P-Value is the best regressions model as far as RMSE is concerned. This is supported by the results in Table 4 where the classifying parameter RSEP is the smallest for this PLS model. The detection of nonlinearity forces one to consider nonlinear models. For instance, one notices that both ANN and RVM yield good performance for the first derivative applied to 10 wavelengths having the smallest Individual P-Value. Thus, FD correction is the best pretreatment for these two nonlinear regressions. This work thus confirms the result that ANN is good at FD correction as reported [29]. Figure 5 is used as an indication of the linearity profile of the NIR-chemometric methods for the paracetamol samples.

## 5. Conclusion

Sixty samples of paracetamol with different batch numbers are scanned in the reflectance mode of a near-infrared spectrometer in the range of 900 nm and 1700 nm. The spectra thus obtained are pretreated by various techniques. The detection and the removal of two outliers lead to the consideration of the fifty-eight samples. The PLS algorithm is applied to the pretreatments of those samples. It reveals that the best regression model is obtained with the MSC correction followed by the first derivative correction when the best subset of the wavelength is chosen by using the p-value approach. The results also found convincing support for the usefulness of the first derivative correction as advocated [29]. The plots of the different pretreatments are reported elsewhere [30]. This manuscript then encourages the potential application of NIR spectroscopy techniques to local paracetamol samples. Hence, this will result in a huge impact on the fight against falsified samples of this antipyretic product. The extension of this application to other directions is of capital interest. As for future work, the extension of the present work to some very demanded drugs should be carried out. Classification of drugs according to batch number is also a possible direction. NIRS techniques which do this classification are for instance principal component analysis (PCA) and the classification using the nonlinear regressions used in this article and [30]. These regression algorithms can also be used to classification.

## Acknowledgements

Aminata Sow recognizes financial support from SIDA (the Swedish International Development Cooperation Agency) through ISP (the International Science Programme, Uppsala University). She greatly appreciates the fruitful discussions with the members at LOSSA. She also thanks the members of the department of physics at USTTB-FST for their constant support.